\documentclass[aps,prl,twocolumn,superscriptaddress,showpacs]{revtex4}

\usepackage{graphicx}
\usepackage{dcolumn} 
\usepackage{bm}      
\usepackage{color}
\usepackage{float}
\newcommand{\ep}{$e$-$ph$~}
\newcommand{\pdag}{{\phantom\dagger}}
\newcommand{\bk}{{\bf k}}
\newcommand{\bq}{{\bf q}}

\begin{document} 

\title{Enhanced superconductivity due to forward scattering in FeSe thin films on SrTiO$_3$ substrates}

\author{Louk Rademaker}
\affiliation{Kavli Institute for Theoretical Physics, University of California Santa Barbara, California 93106, USA}

\author{Yan Wang}
\affiliation{Department of Physics and Astronomy, University of Tennessee, Knoxville, Tennessee 37996, USA}

\author{Tom Berlijn}
\affiliation{Center for Nanophase Materials Sciences, Oak Ridge National Laboratory, Oak Ridge, Tennessee 37831, USA}
\affiliation{Computer Science and Mathematics Division, Oak Ridge National Laboratory, Oak Ridge, Tennessee 37831, USA}

\author{Steve Johnston}
\affiliation{Department of Physics and Astronomy, University of Tennessee, Knoxville, Tennessee 37996, USA}

\date{\today}
  
\begin{abstract} 
We study the consequences of an electron-phonon ($e$-$ph$) interaction that is
strongly peaked in the forward scattering ($\bq = 0$) direction in a
two-dimensional superconductor using Migdal-Eliashberg theory. We find
that strong forward scattering results in an enhanced $T_c$ that  
is linearly proportional to the strength of the dimensionless \ep 
coupling constant $\lambda_m$ in the weak coupling 
limit. This interaction also produces distinct replica bands in the  
single-particle spectral function, similar to those observed in recent 
angle-resolved photoemission experiments on FeSe
monolayers on SrTiO$_3$ and BaTiO$_3$ substrates. By comparing our 
model to photoemission experiments, we infer an \ep
coupling strength that can provide a 
significant portion of the observed high $T_c$ in these systems. 
\end{abstract}



\maketitle 
{\it Introduction} --- A flurry of scientific activities has been generated by the
discovery of an enhanced superconductivity in FeSe monolayers grown
on SrTiO$_3$ (STO) substrates \cite{Wang2012,Liu2012,He2013,Tan2013,LeeNature2014,
PengNatureComm2014,PengPRL2014,GeNatureMat2014,ZhangCPL2014,ZhangPRB2014,LiuPRB2012,XiangPRB2012,coh,
LiJAP2014,Zheng2013,LiuNatureComm2014,HuangPRL2015,Fan,Lee2015,Miyata2015}. On its own, bulk FeSe has a modest
superconducting transition temperature $T_c \sim 9$ K \cite{Bulk}; however, when a monolayer 
is grown on an STO substrate, $T_c$ is increased dramatically \cite{Wang2012}. Most reported $T_c$ 
values cluster within 55 -- 75 K, close to the 
boiling point of liquid nitrogen (77 K). (A surprisingly high $T_c \sim 107$ K has also 
been reported in {\it in situ} transport measurements 
\cite{GeNatureMat2014}.) This discovery has 
opened a pathway to high-$T_c$ superconductivity through interface engineering, 
which has already produced high-$T_c$'s in systems such as 
FeSe on BaTiO$_3$ (BTO) \cite{PengNatureComm2014} and FeTe$_{1-x}$Se$_x$ on STO \cite{LiPRB2015}.  

Determining the origin of the $T_c$ enhancement in these 
interface systems is critical. At the moment, proposals include  
charge transfer between the substrate and FeSe \cite{Liu2012,He2013,ZhangPRB2014,Miyata2015}, 
electric field \cite{Zheng2013} and  
strain effects due to the substrate \cite{Tan2013,PengNatureComm2014}, and phononic 
related effects such as enhanced \ep coupling in the FeSe layer \cite{Wang2012,coh,XiangPRB2012} 
or across the interface \cite{LeeNature2014,Lee2015}. 
Strong evidence for the latter has been provided 
by a recent angle-resolved photoemission spectroscopy (ARPES) study  
\cite{LeeNature2014}, which observed replica bands in the single-particle 
spectral function of the FeSe monolayer. These replicas are interpreted as 
being produced by coupling between the FeSe $3d$ electrons and an optical oxygen
phonon branch in the STO substrate. Moreover, the replica bands are complete
copies of the corresponding main bands, which implies that
the responsible \ep interaction is strongly peaked in the forward scattering direction
(small momentum transfers). Such momentum dependence is notable because 
it can  enhance superconductivity in most pairing channels  
\cite{JohnstonPRL2012,MD,KulicPRB1994,HuangPRB2003,BulutPRB1996,Santi,Kulic,
KulicReview,AperisPRB2011}. 
As such, this cross-interface coupling provides at the same time a suitable 
mechanism for the $T_c$ enhancement in the FeSe/STO and FeSe/BTO 
systems \cite{LeeNature2014,PengNatureComm2014}.

In this \emph{Letter} we explore this possibility and 
examine the consequences of strong forward scattering in the \ep  
interaction for superconductivity and the spectral properties 
of a two-dimensional system.  
By solving the momentum dependent Eliashberg equations, 
we show that a pronounced forward scattering results in a $T_c$ that 
scales linearly with the 
dimensionless \ep coupling constant $\lambda_m$ (see below) in the weak coupling limit. This is in 
stark contrast to the usual exponential dependence predicted by BCS theory. 
Furthermore, this coupling produces distinct replica structures in the 
spectral function similar to those observed experimentally. 
By comparing our model to experiments \cite{LeeNature2014}, we infer 
a significant \ep contribution to the total  
$T_c$ observed in the FeSe/STO system with a modest 
value of $\lambda_m$.

{\it Formalism} --- To model the 
FeSe monolayer we consider a single-band model for the FeSe electron pockets, 
which includes coupling to an oxygen phonon branch in the STO substrate. 
The Hamiltonian is given by 
\begin{eqnarray*}
H&=&\sum_{\bk,\sigma} \xi^{\phantom\dagger}_\bk c^\dagger_{\bk,\sigma}c^{\phantom\dagger}_{\bk,\sigma} 
+ \sum_{\bq} \Omega^{\phantom\dagger}_\bq b^\dagger_{\bq}b^{\phantom\dagger}_\bq 
\\&&+ \frac{1}{\sqrt{N}}\sum_{\bk,\bq,\sigma} g(\bk,\bq)c^\dagger_{\bk+\bq,\sigma}
c^{\phantom\dagger}_{\bk,\sigma} 
(b^\dagger_{-\bq} + b^{\phantom\dagger}_\bq),
\end{eqnarray*}
where $c^\dagger_{\bk,\sigma}$ ($c^\pdag_{\bk,\sigma}$) and $b^\dagger_{\bq}$ ($b^\pdag_{\bq}$) 
are electron and phonon creation (annihilation) operators, respectively, $\xi^{\phantom\dagger}_\bk$ is the 
band dispersion, $\Omega^{\phantom\dagger}_\bq$ is the phonon dispersion, 
and $g(\bk,\bq)$ is the momentum-dependent \ep coupling constant.   

We calculate the single-particle self-energy due to the \ep 
interaction using Migdal-Eliashberg theory. Using the Nambu notation with fermionic Matsubara
frequencies $\omega_n = (2n+1) \pi/\beta$, where $\beta = 1/T$ is the inverse temperature, 
the self-energy is $\hat{\Sigma}(\bk,i\omega_n) = 
i\omega_n[1-Z(\bk,i\omega_n)]\hat{\tau}_0 + \chi(\bk,i\omega_n)\hat{\tau}_3 + 
\phi(\bk,i\omega_n)\hat{\tau}_1$, where $\hat{\tau}_i$ are the Pauli matrices, 
$Z(\bk,i\omega_n)$ and $\chi(\bk,i\omega_n)$ renormalize the single-particle mass and 
band dispersion, respectively,  
and $\phi(\bk,i\omega_n)$ is the anomalous self-energy, which is zero in the normal 
state.  
In Migdal-Eliashberg theory, the self-energy is computed by self-consistently evaluating 
the one-loop diagram and is given by 
\begin{eqnarray*}
\hat{\Sigma}(\bk,i\omega_n)&=& 
\frac{-1}{N\beta}\sum_{\bq,m} |g(\bk,\bq)|^2 D^{(0)}(\bq,i\omega_n-i\omega_m) 
 \times \\ 
& &\quad\quad\quad\hat{\tau}_3\hat{G}(\bk+\bq,i\omega_m)\hat{\tau}_3
\end{eqnarray*}
where
$D^{(0)}(\bq,i\omega_\nu) =-\frac{2\Omega^{\phantom\dagger}_\bq}{\Omega^2_\bq + \omega_\nu^2}$
is the bare phonon propagator, and 
$\hat{G}^{-1}(\bk,i\omega_n) = i\omega_n\hat{\tau}_0 - \xi_\bk\hat{\tau}_3 
- \hat{\Sigma}(\bk,i\omega_n)$
is the dressed electron propagator. 
  
In what follows we parameterize the electronic dispersion as 
$\xi_\bk = -2t[\cos(k_xa)+\cos(k_ya)] - \mu$ with $t = 75$ meV and $\mu = -235$ meV. 
This choice in parameters produces at $\Gamma$ an electron-like Fermi pocket with 
$k_F = 0.97/a$ and a Fermi velocity 
$v_F = 0.12$ eV$\cdot a/\hbar$ along the 
$k_y = 0$ line, where $a$ is the in-plane lattice constant. 
This closely resembles the electron pocket at $M$ point measured by ARPES experiment. 
Since first 
principles calculations indicate that the relevant oxygen phonon branch in STO
is relatively dispersionless near the $\Gamma$-point
\cite{Choudhury,LiJAP2014,Wang2015}, we approximate the phonon with a flat
Einstein mode $\Omega^{\phantom\dagger}_\bq=\Omega=100$ meV $(\hbar=1)$, 
which is consistent with the observed energy separation 
of the replica bands \cite{LeeNature2014}. 
Furthermore, as we are interested 
in the case of forward scattering, we neglect any potential fermion momentum 
dependence in the \ep interaction and set $g(\bq) = g_0\exp(-|\bq|/q_0)$,  
as microscopically derived before \cite{LeeNature2014,Lee2015}.
Here, $q_0$ sets the range of the coupling in momentum space. For 
different values of $q_0$ we adjust $g_0$ to obtain the desired value of the 
dimensionless \ep coupling constant $\lambda_m$, which is computed from the 
Fermi surface averaged mass enhancement in the normal state $\lambda_m = \langle
-\frac{\partial \mathrm{Re}\Sigma(\bk,\omega)}{\partial \omega}\big|_{\omega=0}
\rangle$ \cite{Footnote2}. Throughout we 
assume an $s$-wave symmetry for the gap 
function, consistent with the observations of a fully gapped state 
on the Fermi level \cite{Wang2012,LeeNature2014,Fan,PengPRL2014}.
Finally, we neglect the Coulomb pseudopotential 
$\mu^*$.  One can therefore regard our $T_c$ values as upper bounds 
for the \ep contribution to the FeSe/STO system. 

\begin{figure}[t]
 \includegraphics[width=0.8\columnwidth]{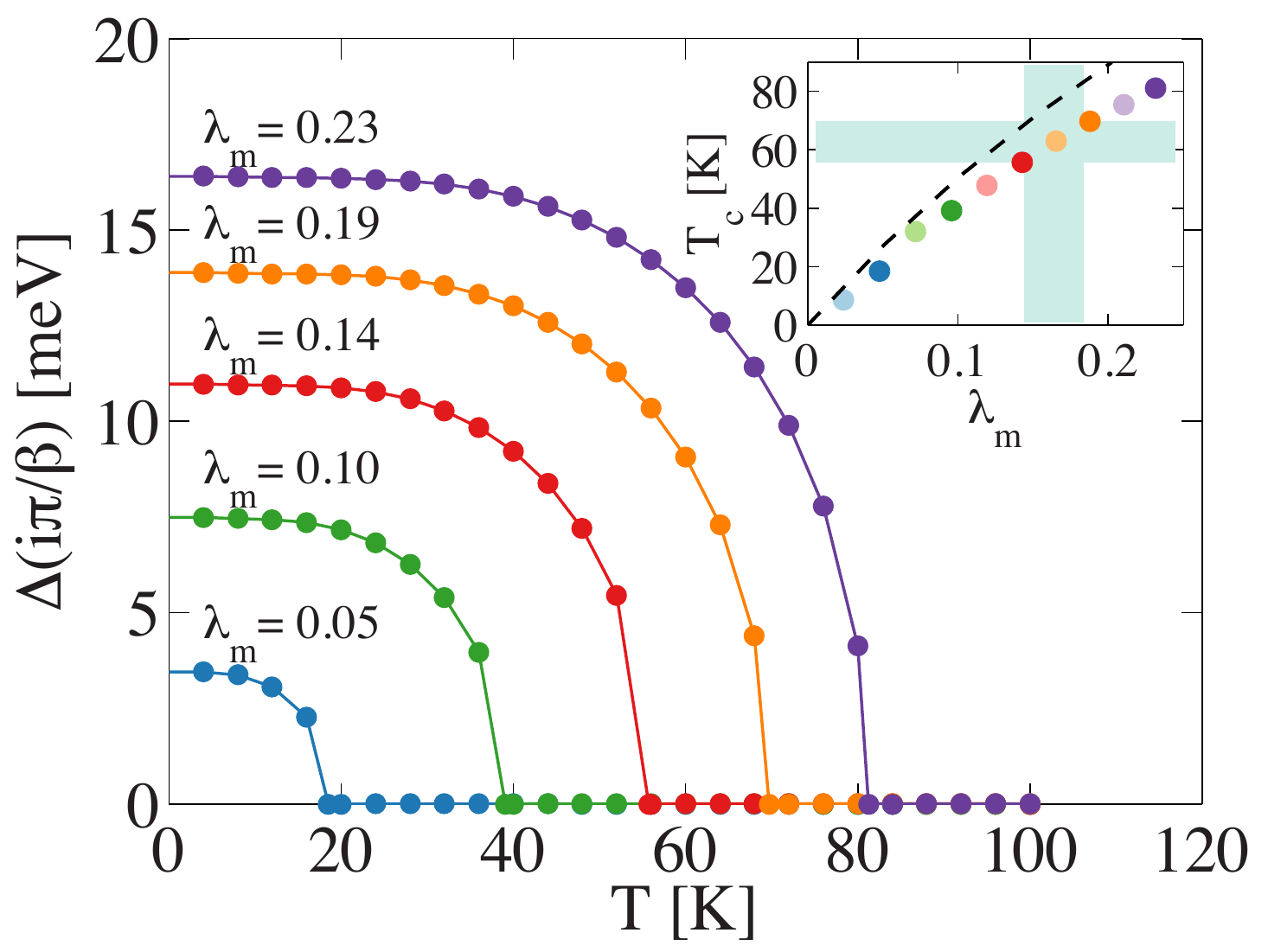}
 \caption{\label{Fig:Tc} (color online) 
 The superconducting gap at the smallest Matsubara frequency 
 $\Delta(i\pi/\beta)$ as a function of temperature for various 
 values of the $e$-$ph$ coupling strength $\lambda_m$, as indicated. 
 The $e$-$ph$ coupling constant $g(\bq)$ is strongly peaked 
 in the forward scattering direction with $q_0 = 0.1/a$. The 
 inset shows $T_c$ as a function of $\lambda_m$, which is 
 extracted from the data in the main panel. The thin 
 dashed line is the result in the limit of perfect 
 forward scattering (see text).
 The shaded area represents the values of $\lambda_m$ that are 
 relevant for FeSe/STO \cite{Supplement}.
}
\end{figure}

{\it Analytical Results} --- 
Before proceeding to full numerical solutions, we can gain some insight by 
first considering the case of perfect forward scattering, 
where the \ep matrix element is a delta function    
$|g(\bq)|^2 = g^2_0\delta_\bq N$ with $g^2_0 = \lambda_m \Omega^2$ \cite{Supplement}. 
In the weak coupling limit, we further set  
$Z(\bk,i\omega_n) = 1$, $\chi(\bk,i\omega_n) = 0$, and therefore 
$\phi(\bk,i\omega_n) = 
\Delta(\bk,i\omega_n)$. With these approximations, the gap function 
on the Fermi surface is given by  
\begin{eqnarray*}\label{Eq:gap}
\Delta(i\omega_n) = \frac{\lambda_m \Omega^2}{\beta}\sum_m 
\frac{\Delta(i\omega_m)}{\omega_m^2 + \Delta^2(i\omega_m)} 
\frac{2\Omega}{\Omega^2 + (\omega_n - \omega_m)^2}.
\end{eqnarray*} 

To determine $T_c$ we take the ansatz 
$\Delta(i\omega_n) = \Delta_0/[1+(\omega_n/\Omega)^2]$ and 
follow the usual steps \cite{Mitrovic}: the gap equation is   
linearlized by setting $\Delta^2_0 = 0$ for $T\sim T_c$ 
and we set $\omega_{n=1}/\Omega = 0$. This results in the condition for $T_c$
\begin{eqnarray*}
1 = \frac{\lambda_m\Omega^2}{\beta_c}\sum_m \frac{2\Omega}
{\omega^2_m(1 + \omega^2_m/\Omega^2)(\Omega^2 + \omega_m^2)}. 
\end{eqnarray*}
The Matsubara sum can be performed exactly, yielding our final expression  
\begin{eqnarray*}\label{Eq:Tc}
1 = \frac{\lambda_m\beta_c}{2}\frac{2\Omega + \Omega\cosh(\Omega\beta_c) 
-(3/\beta_c)\sinh(\Omega\beta_c)}{1+\cosh(\Omega\beta_c)}. 
\end{eqnarray*}

\begin{figure*}[t]
 \includegraphics[width=0.8\textwidth]{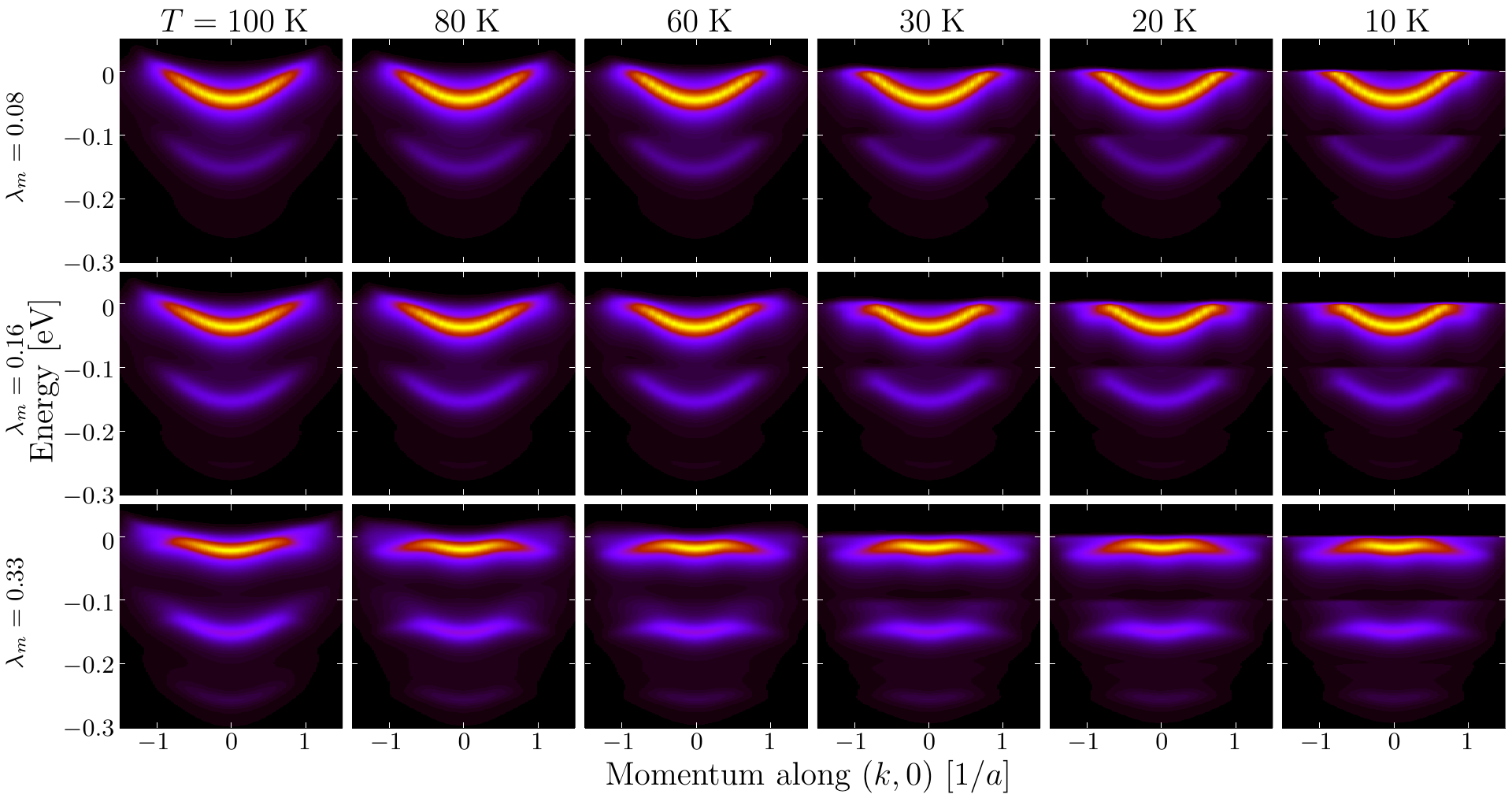}
 \caption{\label{Fig:Spectra} (color online) 
 The temperature dependence of the spectral function 
 for several values of the \ep coupling $\lambda_m$.} 
\end{figure*}

For the case of FeSe, $T_c \ll \Omega$, and the hyperbolic 
functions dominate. To the leading order, the critical temperature is 
\emph{quasi-linear} in the coupling strength in the weak coupling limit,  
$T_c = \frac{\lambda_m}{2+3\lambda_m}\Omega$. 
(A similar result was obtained in Ref. \onlinecite{Kulic} in the context 
of the cuprates using square-well models.) For $\lambda_m = 0.16$ and 
$\Omega = 100$ meV one obtains $T_c = 75$ K, which is 
a remarkably high temperature for such a modest value of $\lambda_m$. 

The increased $T_c$ should be compared to the standard BCS 
value obtained for a momentum-independent
coupling. In this case, the linearized gap equation simplifies 
to \cite{CarbotteRMP,Mitrovic,Supplement}
\begin{eqnarray*}
1 = \pi T_c \lambda_m \sum_{|\omega_m|<\Omega_D}\frac{1}{|\omega_m|} 
= \lambda_m \left[\ln\left(\frac{\Omega_D}{2\pi T_c}\right) - \psi\left(\frac{1}{2}\right)\right],  
\end{eqnarray*}
where we have expanded at large $\Omega_D/T_c$ and $\psi(z)$ is the digamma 
function \cite{Supplement}. This form produces the usual exponential 
behavior for the critical temperature, $T_c = 1.13\Omega_{D}\exp(-1/\lambda_m)$, 
which predicts a $T_c = 2.5$~K for $\lambda_m=0.16$ and $\Omega_{D}=100$ meV.

Comparing these two results, one sees that the origin of the 
enhanced $T_c$ lies in the {\it momentum decoupling} \cite{MD} 
that occurs in the Eliashberg equations when the interaction is strongly 
peaked at $\bq = 0$. In the BCS case, the integration over the Fermi surface  
is equally weighted at all momenta, leading to a $\sum_n \frac{1}{|\omega_n|}$ 
term in the BCS gap equation and subsequently a leading logarithmic behavior.
In the forward scattering case, there is no integration over momentum so
the $\omega_m^{-2}$ term remains, resulting in a leading 
behavior that scales like $1/T_c$ \cite{Supplement}.  
Thus, strong forward scattering serves as an ideal 
mechanism for producing high-$T_c$ superconductivity \cite{KulicReview}. 
Furthermore, a strong forward scattering peak in 
the coupling constant means that this interaction will 
contribute in most pairing channels \cite{LeeNature2014,JohnstonPRL2012,MD,
KulicPRB1994,HuangPRB2003,BulutPRB1996,Santi,Kulic,
KulicReview,AperisPRB2011}. 
It can therefore act in conjunction with other active 
unconventional channels, providing another  
means to increase $T_c$ further.   

{\it Numerical Results for $T_c$ and the superconducting gap} --- 
In real materials the \ep interaction is expected to have a finite range $q_0$ 
in momentum space \cite{LeeNature2014,Wang2015}. Therefore we now consider  
an interaction with a 
finite width by numerically solving the full Eliashberg equations for 
an \ep coupling constant $g(\bq) = g_0\exp(-|\bq|/q_0)$.   
Fig.~\ref{Fig:Tc} shows the superconducting gap at the lowest 
Matsubara frequency $\Delta({\bk_F},i\pi/\beta)$ as a function of temperature for 
several values of $\lambda_m$ and $q_0 = 0.1/a$. We find that the  
superconducting $T_c$ is already large for a modest value of $\lambda_m$ 
and increases approximately linearly with 
$\lambda_m$ in the weak coupling limit; however, the finite range of the coupling 
in momentum space reduces the total $T_c$ slightly with respect to the perfect forward
scattering limit (see the inset of Fig.~\ref{Fig:Tc}). The linear dependence of 
$T_c$ with respect to $\lambda_m$ 
may account for the wide variation of reported $T_c$ values in the literature, 
as differences in sample preparation or doping are likely to result in 
differences in the screening of the \ep coupling and subsequently $T_c$.

{\it Replica Bands} --- The above results show that, in principle, a modest coupling 
to a phonon with a forward scattering peak is capable of accounting for the 
large $T_c$ enhancement observed in FeSe on STO and BTO. The natural question 
is then how much of the experimental $T_c$ is accounted for by this coupling? The 
observed shape and intensity of the replica bands \cite{LeeNature2014,PengNatureComm2014} 
provide us with a direct means to estimate this by comparing our model to experiment. 
To do so, we calculate the single particle spectral function 
$A(\bk,\omega) = -\mathrm{Im}G_{11}(\bk,\omega)/\pi$,  
which requires the analytic continuation of the self-energy to the real 
frequency axis using the method of Ref. \onlinecite{MarsiglioPRB1988}  
(see also \cite{Supplement}).  
Fig.~\ref{Fig:Spectra} plots the temperature evolution of the spectral 
function obtained from a full numerical solution to our model 
for several values of $\lambda_m$, as indicated on the left, and 
$q_0 = 0.1/a$.
In all cases clear replica bands are produced by the coupling, 
offset in energy from the main band by a fixed energy, 
which is $\Omega$ for small values of $\lambda_m$. The separation, however, 
grows for increasing $\lambda_m$.  
This is due to $\chi(\bk,\omega)$, which 
shifts the main band upward in energy. This is most clearly seen 
in the $\lambda_m = 0.33$ results, where the value of $k_F$ has visibly shrunk in the 
main band. In addition, for stronger values of $\lambda_m$ we begin to see the 
formation of a second replica band located at $\sim 2\Omega$ below the main 
band. Thus the observation of only a single replica band in the 
bandstructure of FeSe/STO is consistent with a small $\lambda_m$. 

An intuitive picture for the intensity and energy splitting of the replica band 
can again be obtained in the limit of perfect forward scattering. On the real axis, the 
zero-temperature self-consistent equation for the self-energy in the normal 
state can be written as $\Sigma(\omega) = g^2_0G(\omega + \Omega)$. For 
$\xi_\bk \rightarrow 0^-$, the lowest-order solution is 
$\Sigma(\omega) = \frac{g^2_0}{\omega+\Omega}$ [note that the $\xi_\bk \ne 0$ 
solution can be obtained by shifting the self-energy $\Sigma(\bk,\omega) = 
\Sigma(\omega-\xi_\bk)$]. The poles of the Green's function are at 
$\omega = \Sigma(\omega)$, which has the solution 
$\omega_{\pm} = -\frac{\Omega}{2}\pm \frac{1}{2}\sqrt{\Omega^2 + 4g^2_0}$.
The spectral weight of each pole is given by 
$Z_\pm = [1-\frac{\partial \Sigma}{\partial \omega}|_{\omega = \omega_\pm}]^{-1} 
= [1+\frac{g^2_0}{(\omega_\pm + \Omega)^2}]^{-1}$.
For small $\lambda_m = g^2_0/\Omega^2$, we find that the average 
energy separation between the poles is $\Delta \omega = \Omega [1+2\lambda_m + \mathcal{O}(\lambda_m^2) ]$ 
and the ratio of the spectral weight 
is $\frac{Z_-}{Z_+} = \lambda_m +  \mathcal{O}(\lambda_m^2)$, thus providing 
a direct measure of $\lambda_m$. 

\begin{figure}[tl]
 \includegraphics[width=0.75\columnwidth]{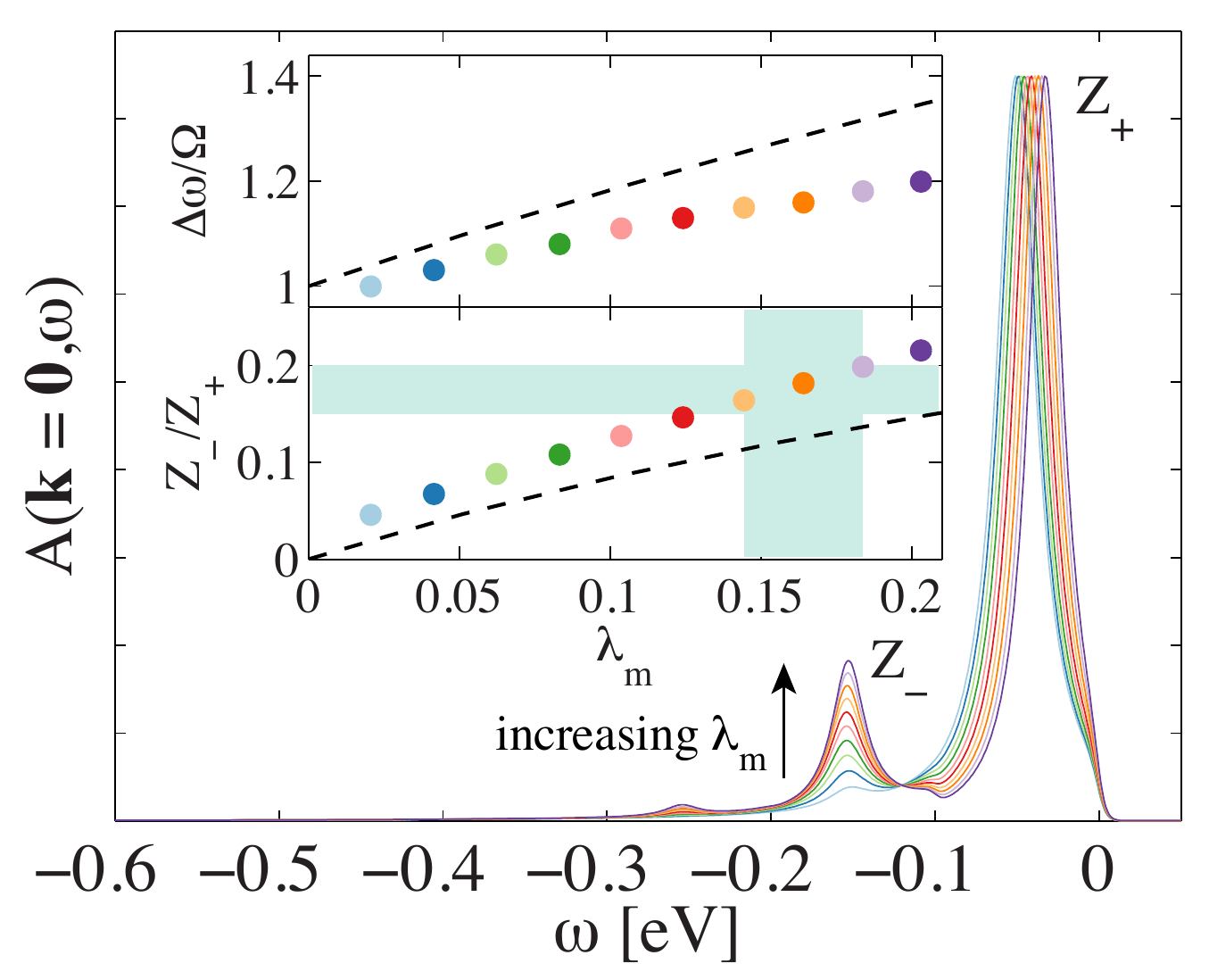}
 \caption{\label{Fig:Intensity} (color online) 
 The spectral function for a momentum at the band bottom ($\bk = 0$ in our model, 
 the $M$ point in the experiment) for $T = 30$ K, $q_0 = 0.1/a$ and  
 $\lambda_m = 0.02 - 0.22$.  The 
 key feature of the forward scattering mechanism is the appearance of the mirror 
 band ($Z_-$) next to the main band ($Z_+$). The relative 
 separation $\Delta\omega$ and intensity $Z_-/Z_+ = A(0,\omega_-)/A(0,\omega_+)$ 
 of these two features  
 is shown in the inset, and increases approximately linearly with $\lambda_m$. 
 The dashed lines show the corresponding result in the perfect forward scattering limit 
 and the shaded area represents the values of $\lambda_m$ that are 
 relevant for FeSe/STO \cite{Supplement}. 
 }
\end{figure}

The spectral weight ratio and energy splitting between the main and replica bands 
can be extracted from our numerical simulations for finite values of $q_0$.  
Fig. \ref{Fig:Intensity} shows $A(\bk,\omega)$ for $\bk = (0,0)$ as a function 
of $\lambda_m$ with $q_0 = 0.1/a$. The behavior matches our expectations gained from 
the perfect forward scattering limit: both the distance between the bands 
and the relative spectral weight grow with increasing $\lambda_m$, though the 
rate of increase is slower than for the case of perfect forward scattering. 
ARPES experiments on the FeSe/STO system \cite{LeeNature2014} observe a spectral 
weight ratio of $\sim 0.15-0.2$ \cite{Supplement}. Comparing to our model 
calculations, we extract a value of $\lambda_m \sim 0.15-0.2$. 
This corresponds to a $T_c \sim 60-70$ K and a gap magnitude of 
$\Delta \sim 10-15$~meV, which are consistent with 
measurements \cite{Wang2012,LeeNature2014,PengPRL2014,Fan}.   


\begin{figure}[tr]
 \includegraphics[width=\columnwidth]{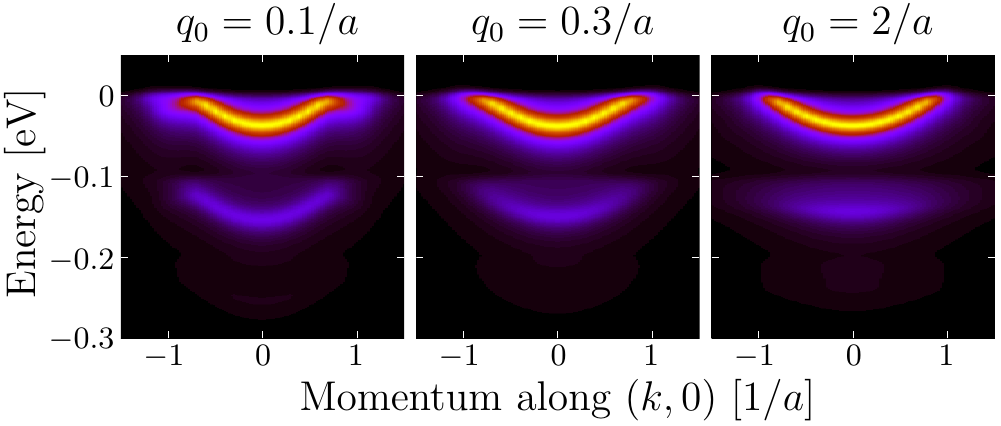}
 \caption{\label{Fig:Akw_vs_q0} (color online) 
 The spectral density $A(\bk,\omega)$ along the $\bk = (k/a,0)$ 
 cut for $q_0 = 0.1$ (left), $0.3$ (middle), and $2$ (right). 
 In all three panels the temperature is $T = 30$ K and $\lambda_m = 0.14$, 
 $0.125$, and $0.25$ in the left, middle, and right panels, respectively.  
 }
\end{figure}

In Fig. \ref{Fig:Akw_vs_q0} we present the evolution of the spectral 
function for increasing values of $q_0$ where $\lambda_m$ is fixed to 
give the same value of $Z_-/Z_+$. 
As expected, the replica bands are observed to smear both in energy and momentum 
as the value of $q_0$ is increased. This shows that a weakly momentum-dependent
coupling (large $q_0$) to an optical mode does not reproduce
the observation of a perfect replica band, with 
the same effective mass and termination points in the Brillouin zone. 
Consequently, strong forward scattering is a necessary ingredient to 
understand the experimental observations \cite{LeeNature2014}.

{\it Summary and Conclusions} --- We have examined the consequences of 
\ep coupling that is strongly peaked in the forward scattering direction 
on the spectral properties and superconducting transition of a two-dimensional 
electronic system.  
We demonstrated that such a coupling produces distinct 
replica bands in the electronic bandstructure consistent with 
recent ARPES measurements on FeSe/STO and FeSe/BTO interface systems. 
In order to reproduce the experimentally observed spectral function, we 
find that relatively modest values of the \ep coupling are needed with 
$\lambda_m \sim 0.15-0.2$. Strong forward scattering results 
in a momentum decoupling of the Eliashberg equations, which subsequently 
produces a larger superconducting $T_c$ in comparison to the predictions 
of conventional BCS theory. As a result, the inferred values of $\lambda_m$ 
predict $T_c$ values on the order of $60-70$~K due to \ep coupling alone. 

We stress that our results do not exclude the presence of another 
unconventional pairing channel such as spin fluctuations. The predicted 
values of $T_c$ and $\Delta$ will be reduced somewhat by the inclusion of the 
Coulomb pseudopotential $\mu^*$. This reduction, however, can be overcome by the 
combination of the \ep and unconventional interactions, since forward 
scattering will contribute to Cooper 
pairing in most channels \cite{LeeNature2014}. An obvious way to distinguish between 
these possible scenarios is to measure the oxygen isotope effect. 
If a purely phononic mechanism is present then $T_c$ should have an isotope 
coefficient $\alpha = -\partial \log(T_c)/\partial \log(M) = 1/2$, while the energy separation 
between the replica bands should decrease by $\sim 0.5(18-16)/16 \sim 6\%$ 
for $^{18}$O rich substrates. Alternatively, 
in a multi-channel scenario, the isotope coefficient $\alpha$ will be reduced from $1/2$ 
when the unconventional channel is significant in comparison to the 
\ep interaction \cite{StevePRB}. This provides a clear means to distinguish 
between these scenarios.

Finally, we note that \ep coupling with a pronounced forward scattering peak has been  
studied in several contexts related to of unconventional superconductivity 
in the cuprates \cite{MD,JohnstonPRL2012,HuangPRB2003,KulicPRB1994,BulutPRB1996,
Santi,KulicReview} and pnictides \cite{AperisPRB2011}. Moreover, it is also 
now being addressed in the context of nematic fluctuations \cite{MaierPRB2014,
LedererPRL2015}. This suggests forward scattering has a 
broader applicability in enhancing superconducting beyond interface systems.

{\it Acknowledgments ---} 
We thank E. Dagotto, T. P. Devereaux, D.-H. Lee, R. G. Moore, D. Scalapino, and J. Zaanen 
for useful discussions. L. R. acknowledges funding from 
Rubicon (Dutch Science Foundation). 
T. B. is supported as a Wigner Fellow at the Oak Ridge 
National Laboratory. A portion of this research was conducted at the Center for
Nanophase Materials Sciences, which is a DOE Office of Science User Facility.
CPU time was provided in 
part by resources supported by the University of Tennessee and Oak 
Ridge National Laboratory Joint Institute for Computational 
Sciences (http://www.jics.utk.edu).


\begin{thebibliography}{99}
\bibitem{Wang2012}
Q. L. Wang {\it et al.},  Chin. Phys. Lett. {\bf 29}, 037402 (2012).
\bibitem{Liu2012}
D. F. Liu {\it et al.}, Nature Commun. {\bf 3}, 931 (2012).
\bibitem{He2013}
S. He {\it et al}., Nature Mater. {\bf 12}, 605–610 (2013).
\bibitem{ZhangPRB2014}
W. H. Zhang {\it et al.}, Phys. Rev. B {\bf 89}, 060506(R) (2014). 
\bibitem{Tan2013}
S. Tan {\it et al.}, Nature Mater. {\bf 12}, 634–640 (2013).
\bibitem{Zheng2013}
F. Zheng, Z. Wang, W. Kang, and P. Zhang, Scientific Reports {\bf 3}, 2213 (2013).
\bibitem{LeeNature2014}
J. J. Lee {\it et al.}, Nature {\bf 515}, 245 (2014).
\bibitem{PengNatureComm2014}
R. Peng {\it et al.}, Nature Commun. {\bf 5}, 5044 (2014).
\bibitem{GeNatureMat2014}
J.-F. Ge, Z.-L. Liu, C. Liu, C.-L. Gao, D. Qian, Q.-K. Xue, Y. Liu, and J.-F. Jia, 
Nature Mater. {\bf 14}, 285 (2014). 
\bibitem{PengPRL2014}
R. Peng {\it et al}., Phys. Rev. Lett. {\bf 112}, 107001 (2014).
\bibitem{ZhangCPL2014}
W. H. Zhang {\it et al.}, Chin. Phys. Lett. {\bf 31}, 017401 (2014).
\bibitem{LiuPRB2012} 
K. Liu, Z.-Y. Lu, and T. Xiang, Phys. Rev. B {\bf 85}, 235123 (2012).
\bibitem{XiangPRB2012} 
Y.-Y. Xiang, F. Wang, D. Wang, Q.-H. Wang, and D.-H. Lee, 
Phys. Rev. B {\bf 86}, 134508 (2012).
\bibitem{LiJAP2014}
B. Li, Z. W. Xing, G. Q. Huang, and D. Y. Xing, Journal of Applied Physics {\bf 115}, 193907 (2014). 
\bibitem{LiuNatureComm2014}
X. Liu {\it et al.}, Nature Commun. {\bf 5}, 5047 (2014).
\bibitem{coh}
S. Coh, M. L. Cohen, S. G. Louie, New J. Phys. {\bf 17}, 073027 (2015).
\bibitem{HuangPRL2015}
D.~Huang {\it et al.}, Phys. Rev. Lett. {\bf 115}, 017002 (2015).
\bibitem{Fan}
Q. Fan {\it et al.}, arXiv:1504.02185 (2015).
\bibitem{Lee2015}
D.-H. Lee, arXiv:1508.02461 (2015).
\bibitem{Miyata2015}
Y. Miyata, K. Nakayama, K. Sugawara, T. Sato, and T. Takahashi, Nat. Mater. \textbf{14} 775 (2015).
\bibitem{Bulk}
F.- C. Hsu, {\it et al.}, Proceedings of the National Academy of Sciences {\bf 105}, 14262 (2008).
\bibitem{LiPRB2015}
F. Li {\it et al.}, Phys. Rev. B {\bf 91}, 220503(R) (2015).
\bibitem{Kulic}
O. V. Danylenko, O. V. Dolgov, M. L. Kuli\'{c}, and V. Oudovenko, 
Eur. Phys. J. B {\bf 9}, 201 (1999).
\bibitem{MD}
G. Varelogiannis, A. Perali, E. Cappelluti, and L. Pietronero, Phys. Rev. B {\bf 54}, R6877 (1996). 
\bibitem{JohnstonPRL2012}
S. Johnston, I. M. Vishik, W. S. Lee, F. Schmitt, S. Uchida, K. Fujita, S. Ishida, 
N. Nagaosa, Z. X. Shen, and T. P. Devereaux, Phys. Rev. Lett. {\bf 108}, 166404 (2012). 
\bibitem{HuangPRB2003}
Z. B. Huang, W. Hanke, and E. Arrigoni, and D.~J.~Scalapino, Phys. Rev. B 68, 220507(R) (2003).
\bibitem{KulicPRB1994}
M. L. Kuli{\'c} and R. Zeyher, Phys. Rev. B {\bf 49}, 4395(R) (1994).
\bibitem{BulutPRB1996}
N. Bulut and D. J. Scalapino, Phys. Rev. B {\bf 54}, 14971 (1996).
\bibitem{Santi}
G. Santi, T. Jarlborg, M. Peter, and M. Weger. Physica C {\bf 259}, 253 (1996). 
\bibitem{KulicReview}
M. L. Kuli{\'c} and O. V. Dolgov, Phys. Stat. Sol. (b) {\bf 242}, 151 (2005) and references therein.
\bibitem{AperisPRB2011}
A. Aperis, P. Kotetes, G. Varelogiannis, and P. M. Oppeneer, Phys. Rev. B {\bf 83}, 092505 (2011). 
\bibitem{Choudhury}
N. Choudhury, E. J. Walter, A. I. Kolesnikov, and C.-K. Loong, Phys. Rev. B {\bf 77}, 134111 (2008). 
\bibitem{Wang2015}
Y. Wang, L. Rademaker, T. Berlijn, and S. Johnston, {\it to be published}.
\bibitem{Supplement}
For further details, please see the online supplementary materials.
\bibitem{CarbotteRMP}
J.~P.~Carbotte, Rev. Mod. Phys. \textbf{62}, 1027 (1990).
\bibitem{Mitrovic}
P. B. Allen and B. Mitrovi{\'c}, in {\it Solid State Physics: Advances in Research and Applications}, 
edited by H. Ehrenreich, F. Seitz, and D. Turnbull. Academic, New York, 1982, Vol. 37, p. 1.
\bibitem{StevePRB}
S. Johnston, F. Vernay, B. Moritz, Z.-X. Shen, N. Nagaosa, J. Zaanen, and 
T. P. Devereaux, Phys. Rev. B \textbf{82}, 064513 (2010).
\bibitem{Footnote2}
We are using $\lambda_m$ to distinguish this definition from the standard 
one involving a double fermi surface average of the coupling constant $|g(\bk,\bq)|^2$. 
See Ref. \onlinecite{Supplement} for further details. 
\bibitem{MarsiglioPRB1988}
F. Marsiglio, M. Schossmann, and J. P. Carbotte, Phys. Rev. B {\bf 37}, 4965 (1988).
\bibitem{LedererPRL2015}
S. Lederer, Y. Schattner, E. Berg, and S. A. Kivelson, Phys. Rev. Lett. {\bf 114}, 097001 (2015).
\bibitem{MaierPRB2014}
T.~A. Maier and D. J. Scalapino, Phys. Rev. B {\bf 90}, 174510 (2014).
\end{thebibliography}
\end{document}


\title{Supplementary Information for \\
Enhanced superconductivity due to forward scattering in FeSe thin films on SrTiO$_3$ substrates}

\author{Louk Rademaker}
\affiliation{Kavli Institute for Theoretical Physics, University of California Santa Barbara, California 93106, USA}

\author{Yan Wang}
\affiliation{Department of Physics and Astronomy, University of Tennessee, Knoxville, Tennessee 37996, USA}

\author{Tom Berlijn}
\affiliation{Center for Nanophase Materials Sciences, Oak Ridge National Laboratory, Oak Ridge, Tennessee 37831, USA}
\affiliation{Computer Science and Mathematics Division, Oak Ridge National Laboratory, Oak Ridge, Tennessee 37831, USA}

\author{Steve Johnston}
\affiliation{Department of Physics and Astronomy, University of Tennessee, Knoxville, Tennessee 37996, USA}

\maketitle

\section{Eliashberg equations}
In this work we have solved the momentum-dependent Eliashberg equations and obtained the electron Green's function
on the real axis, following the method in Ref.~\onlinecite{MarsiglioPRB1988}. Here, we briefly outline the
main aspects of this approach.

\begin{figure}
  \centering
  \includegraphics[width=0.5\columnwidth]{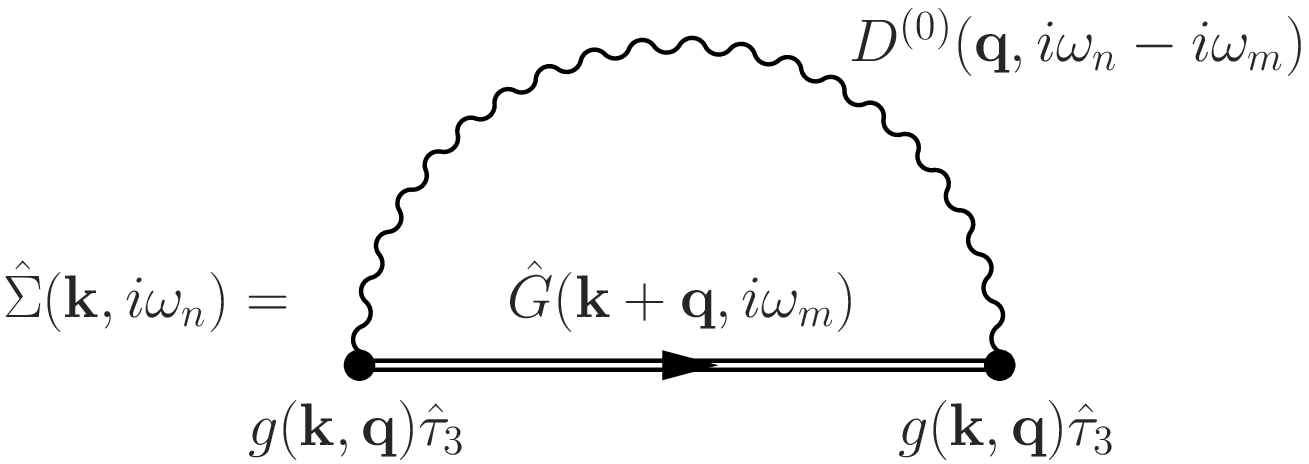}

\caption{The electron self-energy within Midgal-Eliashberg theory is given by the one-loop diagram above,
where the Green's function in the diagram is not the usual bare one but the fully dressed Green's function.
By solving the self-energy self-consistently, we implicitly include higher order terms in the one-loop
diagram, but the vertex corrections are neglected according to Migdal's theorem.}
  \label{Eliashberg}
\end{figure}

In the main text we introduced the electron self-energy in the Eliashberg equation on the imaginary axis with
the Nambu notation
\begin{equation}
  \hat{\Sigma}(\bk,i\omega_n) =
    i\omega_n[1-Z(\bk,i\omega_n)]\hat{\tau}_0 + \chi(\bk,i\omega_n)\hat{\tau}_3 +
    \phi(\bk,i\omega_n)\hat{\tau}_1,
\end{equation}
where $\hat{\tau}_i$ are the usual Pauli matrices, $Z(\bk,i\omega_n)$ and $\chi(\bk,i\omega_n)$ renormalize
the single-particle mass and band dispersion, respectively, and $\phi(\bk,i\omega_n)$ is the anomalous
self-energy, which is zero in the normal state. In Migdal-Eliashberg theory, the self-energy is computed by
self-consistently evaluating the one-loop diagram as shown in Fig.~\ref{Eliashberg} and is given by
\begin{equation}
  \hat{\Sigma}(\bk,i\omega_n) =
    \frac{-1}{N\beta}\sum_{\bq,m} |g(\bk,\bq)|^2 D^{(0)}(\bq,i\omega_n-i\omega_m)
    \hat{\tau}_3\hat{G}(\bk + \bq,i\omega_m)\hat{\tau}_3,
\label{Eli}
\end{equation}
where $D^{(0)}(\bq,i\omega_\nu) = -\frac{2\Omega_\bq}{\Omega^2_\bq + \omega_\nu^2}$ is the bare phonon
propagator, $\hat{G}^{-1}(\bk,i\omega_n) = i\omega_n\hat{\tau}_0 - \xi_\bk\hat{\tau}_3 -
\hat{\Sigma}(\bk,i\omega_n)$ is the dressed electron propagator, $N$ is number of momentum grid points, and
$\beta=1/(k_{\text{B}}T)$ is the inverse temperature. For the dispersionless optical phonon mode we study
here, $\Omega_\bq = \Omega$. For forward scattering the {\ep} matrix element is independent of the fermion
momentum $\bk$ and is given by
\begin{equation}
  |g(\bk,\bq)|^2 = |g(|\bq|)|^2 = g_0^2f(|\bq|),
\end{equation}
where $f(|\bq|)$ is a function peaked at $|\bq|=0$. For the \emph{perfect} forward scattering, we take
$f(|\bq|)=N\delta_\bq$ where $\delta_\bq$ is the Kronecker delta function. For a general forward
scattering we take $f(|\bq|)\propto \exp(-2|\bq|/q_0)$ with a proper normalization factor so that the result
in the $q_0\to 0$ limit can be compared with the result obtained using the Kronecker delta function.

To obtain the electron self-energy for real frequencies, one needs either to perform an analytical
continuation by Pad{\'e} approximants, or to solve the Eliashberg equations directly on the real axis. 
The latter method is 
cumbersome while the former is unreliable; however, Ref.~\onlinecite{MarsiglioPRB1988} introduced a method
whereby the real-frequency self-energy can be found iteratively by using the results on the imaginary axis.
Given the imaginary axis Green's function $\hat{G} (\bk,i \omega_n)$, the real frequency self-energy is found
by solving the iterative equation
\begin{align}
  \hat{\Sigma} (\bk, \omega)
  =& -\frac{1}{N} \sum_\bq \int dz\; |g(\bk,\bq)|^2 B(z) \frac{1}{\beta} \sum_{i \omega_m}
     \frac{ { \hat{\tau}_3} \hat{G} (\bk+\bq,i \omega_m) { \hat{\tau}_3}}{\omega - i \omega_m - z}
	 \nonumber \\
   & +\frac{1}{N} \sum_\bq \int dz\; |g(\bk,\bq)|^2 B(z)
     { \hat{\tau}_3} \hat{G} (\bk+\bq, \omega - z + i\eta) { \hat{\tau}_3}
     \frac{1}{2} \left( \tanh \frac{\omega - z}{2T} + \coth \frac{z}{2T} \right),
	\label{SEcombinedM}
\end{align}
where $B(z) = \delta(z- \Omega) - \delta(z + \Omega)$ is the spectral function of the phonon, and $\eta>0$ is
a small number.

For a specific momentum dependence $f(|\bq|)$ in the coupling function $|g(\bk,\bq)|^2=g_0^2f(|\bq|)$, we set
the coefficient $g_0$ to fix the total dimensionless coupling constant $\lambda_m$, formally defined as (see,
for example, Ref.~\onlinecite{Mahan:2000wd}),
\begin{align}
  \lambda_\bk & \equiv - \text{Re} \left. \frac{\partial \Sigma (\bk,\omega)}{\partial \omega} \right|_{\omega=0}
    = \text{Re} \left[ Z(\bk, \omega) - 1 \right]_{\omega=0}, \\
  \lambda_m   & = \left\langle \lambda_\bk \right\rangle
    = \frac{\sum_\bk \delta(\xi_\bk) \lambda_\bk}{ \sum_\bk \delta(\xi_\bk)}.
	\label{LamDef}
\end{align}
For the relevant low temperature range $T_c\lesssim T \ll \Omega$, the temperature dependencies of the
self-energy in the normal state and $\lambda_m$ are negligible, so we calculate $\lambda_m$ at $T=100$~K, if
not otherwise specified. Here, $\langle\ldots \rangle$ denotes the average over the Fermi surface, and
$\delta(\xi_\bq)$ is approximated by a Gaussian
\begin{equation}
	\delta(x) = \frac{1}{\sqrt{2\pi}\eta} e^{\frac{-x^2}{2\eta^2}}.
\end{equation}
$\lambda_m$ is a measure of the mass enhancement due to electron-phonon coupling, $m^*/m = 1 + \lambda_m$,
where $m^*$ is the effective electron mass.

For a normal metal (that is, with a smooth coupling function $g(\bq)$ and $\Omega\ll E_\text{F}$),
Eq.~(\ref{SEcombinedM})--(\ref{LamDef}) gives the textbook result for the total electron-phonon coupling
constant $\lambda$
\begin{equation}
   \lambda_m \approx \lambda \equiv \frac{2}{N_\text{F}\Omega N^2}
            \sum_{\bk,\bk'}|g(\bk-\bk')|^2 \delta(\xi_\bk)\delta(\xi_{\bk'}),
  \label{NormalLambda}
\end{equation}
where $N_\text{F}=\frac{1}{N}\sum_{\bk}\delta(\xi_\bk)$ is the density of states per spin. For the forward
scattering case, however, the mass enhancement $\lambda_m$ following Eq.~(\ref{LamDef}) differs significantly
from the $\lambda$ defined in Eq.~(\ref{NormalLambda}) for a normal metal. For the forward scattering with a
width $q_0$ in momentum space, $\lambda_m\sim q_0\lambda$ as $q_0\ll k_\text{F}$. For one set of parameters
used in our numerical simulation, $\eta = 15$~meV, $N=64\times64$ discrete momentum points and $q_0=0.1/a$,
we obtain $\lambda_m \approx 0.24 \lambda$ from numerical calculation. The Eq.~(\ref{NormalLambda}) is
therefore unsuitable to characterize the strength of electron-phonon coupling for forward scattering and to
predict the corresponding $T_c$ due to electron-phonon interaction. Consequentially, we use $\lambda_m$ as
the strength of electron-phonon coupling in the main text.

Numerically, we solve the Eliashberg equations with a momentum grid of $N=64 \times 64$ grid points, a
cut-off energy of $600$~meV for the Matsubara and real frequencies with the real frequency discretized into
$1$~meV intervals, and a small-scale parameter $\eta = 15$~meV. For a given set of parameters, we start the
computation with initial self-energies $Z = 1$, $\chi = 0$ and a small anomalous self-energy $\phi =
0.7$~meV. The Eliashberg equations are then iterated, where upon each iteration the chemical potential is
adjusted as to fix the total number of electrons. When the difference between two consecutive self-energies
is less than $10^{-3}$~meV, we have reached the convergence.

\section{BCS gap equation}
For comparison, we derive the standard BCS gap equation from the Eliashberg theory. An extended review of
Eliashberg theory and the BCS limit can be found, for example, in Ref.~\onlinecite{CarbotteRMP}.

In the weak coupling limit, we can neglect the normal electron self-energy by setting $Z(\bk,i\omega_n) = 1$
and $\chi(\bk, i\omega_n)= 0$. The anomalous self-energy is now equal to the gap function,
$\phi(\bk,i\omega_n) = \Delta (\bk,i\omega_n)$. For an isotropic $s$-wave gap, the gap function is
momentum-independent. Under these assumptions the Eliashberg equation reduces to
\begin{eqnarray}
  \Delta (i \omega_n) =
	\frac{-1}{N\beta}\sum_{\bq,m} |g(\bk,\bq)|^2 D^{(0)}(\bq,i\omega_n-i\omega_m)
	\frac{\Delta (i \omega_m)}{\omega_m^2 + \xi_{\bk + \bq}^2 + \Delta^2 (i \omega_m)}.
\end{eqnarray}
Unlike for the forward scattering, we now assume that $|g(\bk,\bq)|^2 D^{(0)}(\bq,i\omega_n-i\omega_m)$ is
momentum independent. Now the only momentum-dependent term in the Eliashberg equation is the electron
dispersion $\xi_{\bk+\bq}$. In this case, the sum over momenta can be replaced by an integral over the free
electron density of states $N(\epsilon)$,
\begin{equation}
	\frac{1}{N} \sum_{\bq} \frac{1}{\omega_m^2 + \xi_{\bq}^2 + \Delta^2 (i \omega_m)}
	=
	\int d\epsilon \, N(\epsilon) \, \frac{1}{\omega_m^2 + \epsilon^2 + \Delta^2 (i \omega_m)}.
\end{equation}
Since most of the relevant physics happens close to the Fermi level, we replace the $N(\epsilon)$ by its
approximate value $N(0)$. The resulting integration over electron energies changes the $1/\omega_m^2$
behavior into a $1/|\omega_m|$,
\begin{equation}
	\int d\epsilon \, N(\epsilon) \, \frac{1}{\omega_m^2 + \epsilon^2 + \Delta^2 (i \omega_m)}
	\approx
	\pi N(0) \frac{1}{\sqrt{\omega_m^2 + \Delta^2 (i \omega_m) }}.	
\end{equation}
The coupling strength as a function of energy can be expressed as $\lambda(i \omega_\nu) = - N(0) |g|^2
D^{(0)}(i\omega_\nu)$ so that we recover the BCS gap equation,
\begin{equation}
	\Delta ( i \omega_n)  =  \pi T \sum_{m} \lambda(i \omega_n - i \omega_m)
	\frac{\Delta (i \omega_m)}{\sqrt{\omega_m^2 + \Delta^2 (i \omega_m)}}
	\label{BCS}
\end{equation}
The critical temperature can be found by taking the limit $\omega_n = 0$ and assuming that
$\lambda(i\omega_n-i\omega_m)$ and $\Delta(i \omega_m)$ are constants $\lambda$ and $\Delta_0$ with a hard
cut-off at the Debye frequency $\Omega_D$. The linearized gap equation with $\Delta_0^2 = 0$ now takes the
form
\begin{equation}
  1  =  \pi T_c \lambda \sum_{|\omega_m| < \Omega_D} \frac{1}{|\omega_m|}
  \approx \lambda \left[ \ln \left( \frac{\Omega_D}{2 \pi T_c} \right) - \psi \left(\frac{1}{2} \right) \right],
  \label{LinGapBCS}
\end{equation}
where we have expanded at large $\Omega_D/T_c$ and $\psi(z)$ is the digamma function. The critical
temperature is given by the exponential function of the total coupling constant $\lambda$
\begin{equation}
	T_c = 1.13 \Omega_D e^{-1/\lambda}.
\end{equation}
This exponential suppression can be traced back to the integration over the electron energies. This, in turn,
is possible because the vertex and phonon part of the gap equation are momentum independent. When the vertex
has strong momentum-dependence, the integration over electron energies cannot be performed and the leading
behavior is still $1/\omega_m^2$ instead of $1/|\omega_m|$, as will be shown when we consider perfect forward
scattering.

\section{Perfect forward scattering}
For the forward scattering with $g(\bq) \propto g_0\exp(-|\bq|/q_0)$, the Eliashberg equations can be solved
analytically in the limit of \emph{perfect} forward scattering, $q_0 \to 0$. The results of this limit are
discussed in the main text; here we provide a more detailed derivation of the proper coupling strength
$\lambda_m$ and the corresponding critical temperature $T_c$.

The coupling for the perfect forward scattering is
\begin{equation}
	|g (\bq)|^2 = g_0^2 N\delta_\bq
\end{equation}
on a discrete lattice with total $N$ grid points, where $N\delta_\bq$ is the Kronecker delta function
normalized to unity in the sum $\frac{1}{N}\sum_\bq N\delta_\bq = 1$. (In the large $N$ and continuum limit,
$N\delta_\bq \to (2\pi)^2\delta(\bq)$, and the Dirac delta function $\delta(\bq)$ can be simulated with a
peak function with finite width $q_0$, as for the general forward scattering case.) The one-loop electron
self-energy now reads
\begin{eqnarray}
  \Sigma^{(1)}(\bk,\omega) = \frac{g_0^2}{2} \int dz \frac{B(z)}{ \omega - \xi_{\bk} - z + i \eta}
    \left( \tanh \frac{\xi_{k}}{2T} + \coth \frac{z}{2T} \right),
\end{eqnarray}
where $B(z) = \delta( z - \Omega) - \delta(z + \Omega)$. The subscript ``(1)'' of the self-energy means we 
have performed a single iteration of the solving the Eliashberg equation, where we 
substitute in the bare electron Green's function instead of the dressed Green's function. At the
Fermi surface, the electron energy vanishes ($\xi_\bk = 0$) so the self-energy simplifies to
\begin{equation}
  \text{Re}\,\Sigma^{(1)} (\bk_\text{F}, \omega) =
    g_0^2 \coth \left( \frac{\Omega}{2T} \right) \frac{\omega}{\omega^2 - \Omega^2},
\end{equation}
which is momentum independent. In the low temperature limit, $\coth(\Omega/2T)\approx 1$, and by taking the
derivative of the self-energy at $\omega =0$ we obtain the total coupling constant as defined in
Eq.~(\ref{LamDef})
\begin{equation}
	\lambda_m = - \left\langle \text{Re}
      \left. \frac{\partial \Sigma^{(1)}(\bk_F,\omega)}{\partial \omega} \right|_{\omega=0}
      \right\rangle = \frac{g_0^2}{\Omega^2}.
\end{equation}
The numerical value of $\lambda_m$ from the self-consistently calculated self-energy is very close to the
value given by the above equation. For a nearly perfect forward scattering with a peak width $q_0=0.1/a$,
even after considering the small temperature dependence, the numerical value differs from the above result by
a few percent.

We now continue to derive the linear dependence of transition temperature $T_c$ in terms of the coupling
strength $\lambda_m$ in the weak coupling limit for the perfect forward scattering. In this limit, we can
neglect the normal electron self-energies, that is $Z (\bk,i \omega_n) = 1$ and $\chi (\bk,i \omega_n)=0$. We
assume an isotropic $s$-wave gap. After summing over the momentum $\bq$,  we obtain the self-consistent
equation for the gap function $\phi(\bk,i \omega_n) = \Delta(\bk,i \omega_n)$,
\begin{equation}
  \Delta (i \omega_n) = \frac{\lambda_m \Omega^2}{\beta} \sum_{\omega_m}
	\frac{\Delta( i \omega_m)}{\omega_m^2 + \Delta^2(i \omega_m)}
	\frac{2 \Omega}{\Omega^2 + (\omega_n - \omega_m)^2}.
  \label{SCEqnPFS}
\end{equation}
To find the critical temperature, we take an ansatz for the gap function with the form
\begin{equation}
  \Delta(i\omega_n) = \frac{\Delta_0}{1 + (\omega_n/\Omega)^2}.
\end{equation}
Our numerical results verify that this shape properly represents the Matsubara frequency dependence of the
gap. Setting $\omega_{n=1}/\Omega=0$ and linearizing the gap equation with $\Delta_0^2=0$, we find the
equation for the critical temperature,
\begin{equation}
  1 = \frac{\lambda_m \Omega^2}{\beta_c} \sum_{\omega_m}
      \frac{2 \Omega}{\omega_m^2[1 + (\omega_m/\Omega)^2](\Omega^2 + \omega_m^2)}
  \label{LinGapFS}
\end{equation}
where the sum over Matsubara frequencies can be performed exactly by the method of contour integration. We
finally obtain
\begin{equation}
  1 = \frac{\lambda_m}{2T_c}
      \frac{ 2\Omega + \Omega\cosh\frac{\Omega}{T_c} - 3T_c\sinh\frac{\Omega}{T_c} }
           { 1 + \cosh\frac{\Omega}{T_c} }.
\end{equation}
It is reasonable to assume that the critical temperature $T_c$ is much less than the optical phonon
frequency, as $\Omega$ is typically of the order of 100 meV $\approx 1000$ K. In this limit, the hyperbolic
functions dominate and cancel each other in the numerator and denominator of the above equation. We find a
critical temperature of
\begin{equation}
  T_c = \frac{\lambda_m}{2 + 3\lambda_m} \Omega.
  \label{PFStc}
\end{equation}
This gives a remarkably high critical temperature, even for a modest coupling constant $\lambda_m$ because of
the quasi-linear dependence.  For example, $\lambda_m=0.16$ with $\Omega = 100$~meV yields a critical
temperature of $T_c=75$~K.

\section{Spectral weight ratio and $\lambda_m$ from ARPES data}
\begin{figure}
  \centering
  \includegraphics[width=0.5\columnwidth]{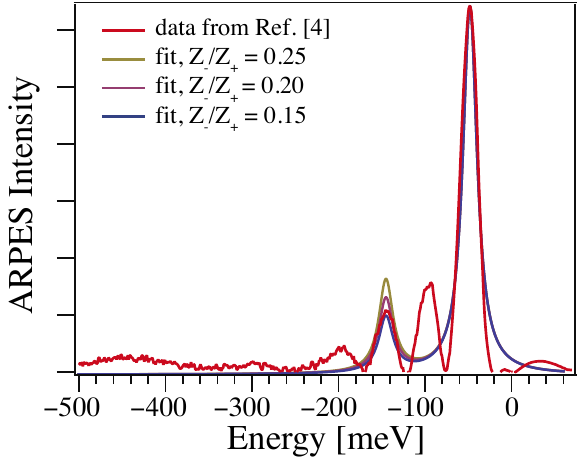}

\caption{ARPES data (integrated and background subtracted) from Lee \textit{et al.}~\cite{LeeNature2014} is
fitted using the sum of two Lorentzians with height ratio 0.25, 0.2 and 0.15. The ratios in the range
$Z_-/Z_+ \sim 0.15\text{--}0.2$ fit the experimental data best. The peak at $-100$~meV energy in ARPES data
is not included in our single band model calculations.}
  \label{FigFit}
\end{figure}

By fitting ARPES data from Lee \textit{et al.}~\cite{LeeNature2014}, we found that the ratio of spectral
weights at the peak of the main band and the mirror band is approximately $Z_-/Z_+ \sim 0.15\text{--}0.2$, as
shown in Fig.~\ref{FigFit}. This implies a value of the coupling constant $\lambda_m \sim 0.15\text{--}0.2$.
